# Magnetoelectric and Raman spectroscopic studies of single-crystalline MnCr$_2$O$_4$

G. T. Lin[1,2,†], Y. Q. Wang[3,2,†], X. Luo[1,*], J. Ma[4,5], H. L. Zhuang[6], D. Qian[5,7], L. H. Yin[1],

F. C. Chen[1,2], J. Yan[1,2], R. R. Zhang[3], S. L. Zhang[3], W. Tong[3], W. H. Song[1],

P. Tong[1], X. B. Zhu[1], and Y. P. Sun[3,1,7*]

[1]*Key Laboratory of Materials Physics, Institute of Solid State Physics,*

*Chinese Academy of Sciences, Hefei 230031, China*

[2]*University of Science and Technology of China, Hefei 230026, China*

[3]*High Magnetic Field Laboratory, Chinese Academy of Sciences,*

*Hefei 230031, China*

[4]*Quantum Condensed Matter Division, Oak Ridge National Laboratory,*

*Oak Ridge, Tennessee 37831, USA*

[5]*Department of Physics and Astronomy, Shanghai Jiao Tong University,*

*Shanghai 200240, China*

[6]*School for Engineering of Matter, Transport & Energy, Arizona State University,*

*Tempe, Arizona 85287, USA*

[7]*Collaborative Innovation Center of Advanced Microstructures,*

*Nanjing University, Nanjing 210093, China*

[†] These authors equal to this work.

[*]Corresponding authors: xluo@issp.ac.cn and ypsun@issp.ac.cn.

**Abstract**

$MnCr_2O_4$ that exhibits spin frustration and complex spiral spin order is of great interest from both fundamental as well as application-oriented perspectives. Unlike $CoCr_2O_4$ whose ground state presents the coexistence of commensurate spiral spin order (CSSO) and ferroelectric order, $MnCr_2O_4$ shows no multiferroicity. One reason is that the spiral spin order is highly sensitive to the oxygen concentration in $MnCr_2O_4$. Here, we have successfully grown high-quality single-crystalline $MnCr_2O_4$ by the chemical vapor transport method. We observe a new first-order magnetic transition from the incommensurate spiral spin order (ICSSO) at 19.4 K to the CSSO at 17.4 K. This magnetic transition is verified by magnetization, specific heat, and magnetoelectric measurements, which also confirm that the ground state exhibits the coexistence of the CSSO and magnetoelectricity below 17.4 K. Interestingly, the temperature evolution of Raman spectra between 5.4 and 300 K suggests that the structure remains the same. We also find that the phase-transition temperature of the CSSO decreases as applied magnetic field increases up to 45 kOe.

# I. Introduction

Insulators with spiral spin order, offering immense potential in low loss memory devices, have attracted significant interest due to their multiferroicity (MF) in which the dielectric and magnetic polarizations can be manipulated by applying either magnetic fields or electric ones [1-9]. In these MFs, spinel compounds with cubic structure are an important class of materials and their electronic properties have drawn high attention due to their colossal magnetocapacitance and spontaneous dielectric polarization in the magnetically ordered state [1,2,10-16]. The appearance of MF is associated with either non-collinear spiral spin order or off-centering of magnetic ions from their symmetric site positions in the lattice [1,2,10-17]. Nevertheless, single-crystalline $MnCr_2O_4$, showing a complex spiral spin order similar to $CoCr_2O_4$, does not show MF effects up to now [1,2,15-30].

$MnCr_2O_4$ crystallizes in a cubic spinel structure with $Fd\bar{3}m$ space group (shown in Figure 5(a)), where magnetic $Mn^{2+}$ ($3d^5$, $S = 5/2$) and $Cr^{3+}$ ($3d^3$, $S = 3/2$) ions occupy the tetrahedral and octahedral sites, respectively. A long-range ferrimagnetic spin order (LFIM) appears below $T_C$ = 41-52 K, followed by transition into a short-range spiral spin order at $T_S$ = 14-20 K [15,18-23,25,26]. Compared with $CoCr_2O_4$, an incommensurate spiral spin order (ICSSO) to commensurate spiral spin order (CSSO) transition remains to be clarified in $MnCr_2O_4$ [1,2,15-23,25-31]. And it is obvious that the spiral spin order may be very sensitive to oxygen content [31], which may be one of the reasons that the magnetic ground state of $MnCr_2O_4$ is as yet in dispute. Therefore, firstly, the high quality single-crystalline $MnCr_2O_4$ is successfully grown by the chemical vapor transport method (CVT). Then, we present a detailed investigation of magnetic ground state in single-crystalline $MnCr_2O_4$. We find that a first-order transition from ICSSO to CSSO with magnetoelectricity (ME) occurs at $T_L$ = 17.4 K, indicating strong spin-lattice coupling. Interestingly, the temperature evolution of Raman spectrum between 5.4 and 300 K indicates that there is no structural phase change in $MnCr_2O_4$.

## II. Experimental and theoretical details

Samples of single-crystalline $MnCr_2O_4$ were grown by the CVT, with $CrCl_3$ powders as the transport agent. Experimental details concerning the preparation of $MnCr_2O_4$ were given in Ref. [19]. The x-ray diffraction (XRD) data indicated that the powders were single phase with cubic structure (see the Supplemental Material [32]). We measured the specific heat (SH) using the Quantum Design physical properties measurement system (PPMS-9T) and characterized the magnetic properties by the magnetic property measurement system (MPMS-XL5). The x-ray photoelectron spectra (XPS) were measured in Thermo ESCALAB 250 spectrometer using Al Kα x-ray at 1486.6eV as the excitation source (see the Supplemental Material [32]). A plate with the (111) plane of 0.5 × 0.5 $mm^2$ was polished from the single crystal. The specimen was cooled down to 2 K with an applied electric field $E$ = 0 V along [111] at different applied magnetic fields parallel to (111) plane. Raman-scattering experiments were conducted by using the 780 nm laser line in a DXR Raman Microscope (Thermo Scientific). The scattering light was collected by using a single exposure of the CCD with a spectral resolution of 1 $cm^{-1}$. Low-temperature Raman spectra were obtained on a Raman Microscope (Horiba JY T64000) equipped with Janis ST-500 microscopy cryostat.

We used the Vienna *Ab initio* Simulation Package [33] to calculate the force constants and Raman spectrum [34] of $MnCr_2O_4$. Plane waves with a cutoff energy of 500 eV were employed to model the valence electrons. We used the potentials based on the projector augmented-wave method [35,36]. We also used the Perdew–Burke–Ernzerhof functional [37] to describe the exchange-correlation interactions. Similar to a previous theoretical study [38], the Coulomb interactions between *d* orbitals of Mn and Cr atoms were treated with Dudarev's effective *U-J* parameters [39] of 3.0 and 5.0 eV, respectively. We computed the force constants using $3 \times 3 \times 3$ supercells and these force constants were post-processed by the PHONOPY program [40] to obtain the phonon spectrum of $MnCr_2O_4$. *K*-point meshes based on the Monkhorst-Pack scheme [41] were $6 \times 6 \times 6$ and $2 \times 2 \times 2$ for the unit cell and supercells, respectively.

## III. Results and discussions

Figure 1(a) shows the temperature-dependent magnetization ***M***(*T*) of $MnCr_2O_4$ under the zero-field-cooled warming (ZFC), field-cooled cooling (FCC) and field-cooled warming (FCW) modes with applied magnetic field *H* = 50 Oe, parallel to the <111> direction. We observe a paramagnetic–ferrimagnetic (PM-FIM) transition that occurs at $T_C$ of 40 K, as determined by the derivative of the magnetization. This temperature is close to the values of 41-52 K reported previously [15,18-23,25,26]. For a FIM system, the temperature-dependent inverse susceptibility $\chi(T)^{-1}$ above $T_C$ can be described by the hyperbolic behavior characteristic of ferrimagnets resulting from the mean-field theory [25,42].

$$\frac{1}{\chi} = \frac{T-\theta}{C} - \frac{\zeta}{T-\theta'}, \qquad (1)$$

where $C$ is the Curie constant, $\theta$ is the Weiss temperature, the first term is the hyperbolic high-$T$ asymptote that has a Curie-Weiss (CW) form and the second term is the hyperbolic low-$T$ asymptote. The fitted $\chi(T)^{-1}$ for MnCr$_2$O$_4$ by Eq. (1) is shown by the red curve in the inset of Figure 1(a) using the $C$ = 8.51 emu·K/mol, $\theta$ = -410.9 K, $\zeta$ = 1449.2 mol·K/emu, and $\theta'$ = 20.6 K. The effective magnetic moment is determined to be $\mu_{eff}$ ~ 8.25 $\mu_B$ ($\mu_{eff} = \sqrt{3k_B C / N_A}\mu_B$), which is close to the theoretical value expected for high spin Cr$^{3+}$ (S = 3/2) and Mn$^{2+}$ (S = 5/2) [25]. The high ratio of $|\theta|/T_C$ ~ 10 indicates significant magnetic frustration due to competing $J_{CrCr}$, $J_{MnMn}$, and $J_{MnCr}$ exchange interactions that establish the spiral spin order at low temperatures [21,22,25]. Figure 1(b) shows the isothermal magnetization $M(H)$ at 5 K and the $M(H)$ curve presents almost no coercive force for single-crystalline MnCr$_2$O$_4$. Above about 3 kOe, the magnetization increases linearly with applied field exhibiting a typical FIM behavior (parallel to <111> direction). In the insets (i) and (ii) of Figure 1(b), time-dependent magnetization $M(t)$ of MnCr$_2$O$_4$ under FCC mode is recorded below $T_C$. The sample is dropped to the desired temperatures from well above $T_C$ in 100 Oe and decay of $M(t)$ is recorded with $t$. The $M(t)$ is fitted with the modified stretched exponential function [15]

$$M(t) = M_0 - M_g \exp\left[-(t/\tau)^\beta\right], \qquad (2)$$

where, the $M_0$ and $M_g$ are the ferromagnetic and exponential components of $M(t)$, respectively. $\beta$ is an exponent with the range $0 < \beta < 1$. The fitted curve using the Eq.

(2) is displayed in the inset of Figure 1(b). The values of $M_g/M_0$, $\tau$, and $\beta$ are 90.3%, 2004 s, and 0.58, respectively, at 3 K and 88.7%, 972 s, and 0.64, respectively, at 18 K. $\beta < 1$ indicates the relaxation mechanism with the spin-glass-like behavior (SG-like) [15]. Compared to the $M(t)$ with no the relaxation above $T_S = 19.4$ K, the significant relaxation of $M(t)$ is observed below $T_S$, which does not agree with the previous reports that the relaxation with the SG-like can be also seen between $T_C$ and $T_S$ [15,18-21]. One of the reasons for the difference is attributable to the defects caused by oxygen vacancies in the different $MnCr_2O_4$ samples, which can be confirmed by the inset (ii) of Fig. 1(a). Here, we compare the two selected samples named the Sample 1 (this work) and Sample 2 (single-crystalline $MnCr_2O_4$ of Ref. [18,19]). We observe a PM-FIM transition that occurs at $T_C$ of 39.5 K (Sample 1) and 52 K (Sample 2), as determined by the derivative of the magnetization. For Sample 2, one can see the phenomena as follow: the anomaly at $T_S \approx 18$ K in the FCC curve, a weak thermal hysteresis observed in the FC magnetization, and reentrant-spin-glass-like characteristic temperature $T_t \approx 47$ K, which is consistent with the previous reports [15,18,19,21]. In contrast to the Sample 2, there are a lower PM-FIM transition temperature $T_C$, an anomaly independent on the applied magnetic field at $T_S$, and a first-order transition at $T_L$ for the Sample 1 (More details see the Supplemental Material [32]).

In order to show clearly the low-temperature phase transitions, we plot the enlarged view of the $M(T)$ under FCC and FCW modes with different applied magnetic field, parallel to the (111) plane (Figure 1(c)-1(e)). By analogy with

CoCr$_2$O$_4$ [17,21,27,29], it is noticed that all $M(T)$ curves show a steplike kink at the ICSSO transition at $T_S$ = 19.4 K. If the temperature further decreases, the $M(T)$ curves exhibit a second anomaly at the transition into the CSSO at $T_L$ = 17.4 K with a hysteresis under FCC and FCW modes, confirming the first-order magnetic phase transition. With an increasing magnetic field, the anomaly in the magnetization at $T_L$ shifts to lower temperatures, whereas the position of the anomaly at $T_S$ remains unchanged.

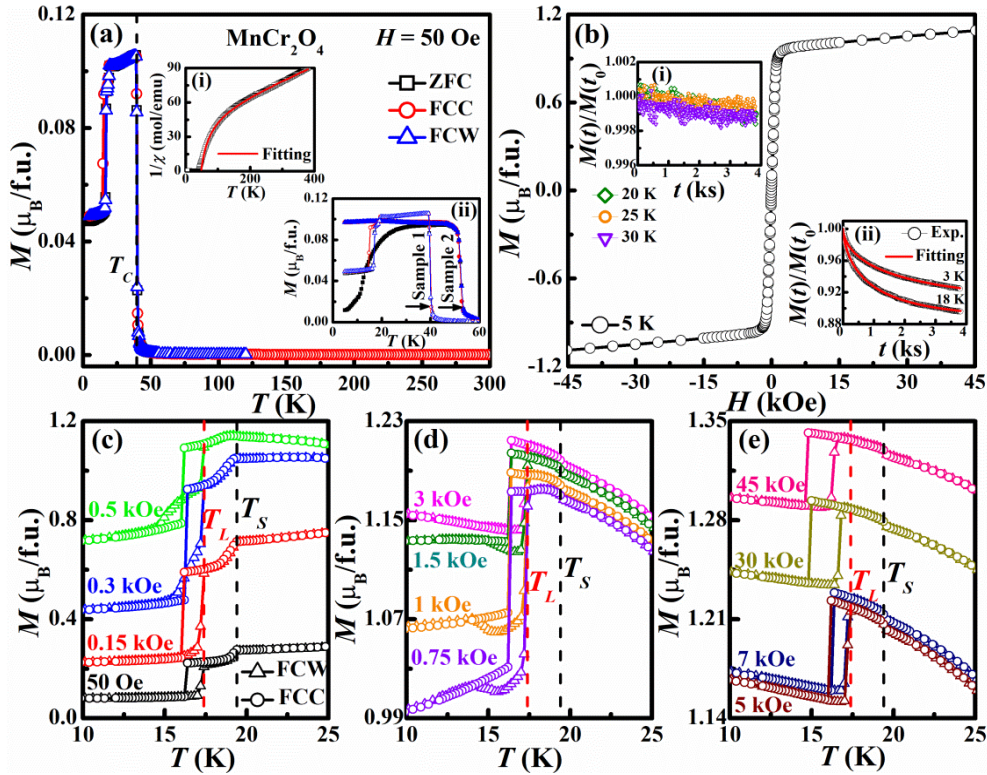

Figure 1. (Color online): (a) Temperature-dependent magnetization $M(T)$ of MnCr$_2$O$_4$ under the ZFC, FCC, and FCW with an applied magnetic field $H$ of 50 Oe, parallel to the <111> direction. The inset (i) shows the temperature-dependent inverse susceptibility $\chi(T)^{-1}$. The red solid line is the fitting result according to Eq. (1). The inset (ii) shows the temperature-dependent magnetization $M(T)$ measurements of the sam.1 and 2 with $H$ = 50 Oe, parallel to the <111> direction; (b) The isothermal magnetization curves $M(H)$ at 5 K, parallel to the <111> direction. The insets (i) and (ii) present the time-dependent magnetization $M(t)$ under FCC mode below $T_C$; (c), (d) and (e) The enlarged view of the $M(T)$ under FCC and FCW modes with different applied magnetic field, parallel to the (111) plane.

Figure 2(a) shows the variation of the zero-field SH $C_p(T)$ with temperature under cooling mode. The sharp anomaly in $C_p(T)$ at $T_C^{SH}$ = 40 K corresponds to the FIM transition temperature, followed by a transition into an ICSSO at $T_S^{SH}$ = 19.5 K. The most striking feature with a sharp peak at $T_L^{SH}$ = 17.5 K is ascribed to the CSSO. Since MnCr$_2$O$_4$ is insulator, the electronic contribution to the heat capacity is not considered. The $C_{mag}$ can be calculated by the following equations [42]:

$$C_{mag}(T) = C_p(T) - nC_V^{Debye}(T), \qquad (3)$$

and

$$C_V^{Debye}(T) = 9R\left(\frac{T}{\Theta_D}\right)^3 \int_0^{\Theta_D/T} \frac{x^4 e^x}{(e^x-1)^2} dx, \qquad (4)$$

where $R$ is the molar gas constant, $\Theta_D$ is the Debye temperature, and $n = 7$ is the number of atoms per formula unit. The sum of Debye functions accounts for the lattice contribution to SH. We can extract the magnetic contribution $C_{mag}(T)$ from the measured SH of MnCr$_2$O$_4$. The fitted $C_p(T)$ by Eq. (3) and (4) over the temperature range from about 3 to 200 K is shown by the red curve in Figure 2(a) using the Debye temperature $\Theta_D$ = 750 K. This value compares well to the Debye temperature in ferrites [43]. The $C_{mag}(T)$ curve exhibits three clear features, indicative of three phase transitions, as displayed in Figure 2(b). The magnetic entropy $S_{mag}(T)$ is calculated by

$$S_{mag}(T) = \int_0^T \frac{C_{mag}(T)}{T} dT. \qquad (5)$$

The inset of Figure 2(b) shows the temperature dependence of $S_{mag}(T)$. The entropy of MnCr$_2$O$_4$ per mole with completely disordered spins S is

$$S_{mag}(T \to \infty) = R\ln(2S_{Mn^{2+}}+1) + 2R\ln(2S_{Cr^{3+}}+1). \qquad (6)$$

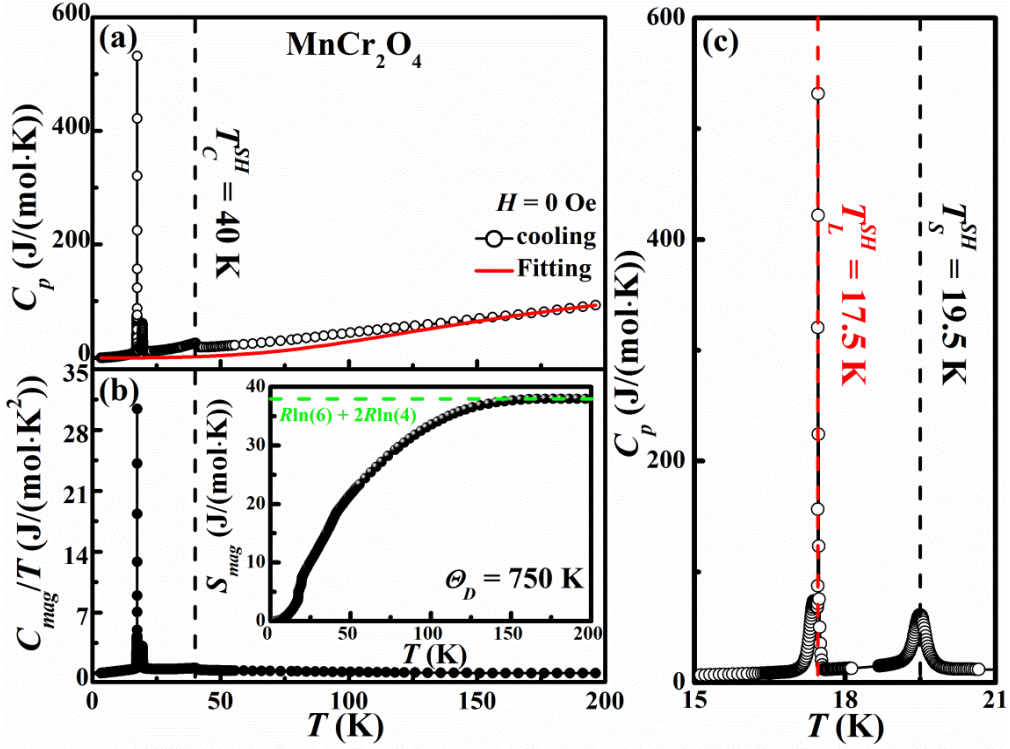

Figure 2. (Color online): (a) Specific heat $C_p$ as a function of $T$ for MnCr$_2$O$_4$ and the fitted $C_V^{Debye}(T)$ using Eq. (3) and (4); (b) Temperature-dependent magnetic specific heat $C_{mag}(T)$. The inset shows the magnetic entropy $S_{mag}(T)$. The red dashed line refers to $S_{mag}(T \to \infty)$ calculated with the magnetic moment S = 5/2 for Mn$^{2+}$ and S = 3/2 for Cr$^{3+}$; (c) The enlarged view of the $C_p(T)$ under cooling and warming modes with $H$ = 0 Oe.

Using S = 5/2 for Mn$^{2+}$ and S = 3/2 for Cr$^{3+}$, we obtain $S_{mag}(T \to \infty)$ of 37.9 J/(mol·K). However, we observe the $S_{mag}$ is 17.8 J/(mol·K) at $T_C^{SH}$, which is only 47% of the value excepted for $S_{mag}(T \to \infty)$. Note that there is an error of about 10% [44] in our measurement due to the fitting of the optical phonon contributions at high temperatures. In spite of this small error, our result indicates the strong dynamic short-range spin interactions above $T_C$. In addition, we plot the enlarged view of the $C_p(T)$, as shown in Figure 2(c). There are two extremely sharp peaks at $T_S^{SH}$ and $T_L^{SH}$, indicating the first-order transition (see the Supplemental Material [32]).

To determine if there is a structural transition near the magnetic transitions in MnCr$_2$O$_4$, the temperature evolution of Raman spectrum in the energy range 150 -

720 cm$^{-1}$ is studied in details between 5.4 and 300 K, as shown in Figure 3(a). MnCr$_2$O$_4$ has a cubic ($Fd\bar{3}m$) structure with five Raman active modes which are classified as $\Gamma_{Raman}$ = 3T$_{2g}$ + A$_{1g}$ + E$_g$ [45]. The spectra in Figure 3(a) show no indication of the splitting of the T$_{2g}$ and A$_{1g}$ between 5.4 and 300 K at least to within the resolution of this experiment, indicating no structure change. This result is in agreement with diffraction studies and optical conductivity spectra of MnCr$_2$O$_4$ [15,21,26]. And we displays the calculated phonon and Raman spectra of MnCr$_2$O$_4$ (see the Supplemental Material [32]). One can observe the same Raman-active modes as seen in the experiment. In particular, the A$_{1g}$ mode exhibits the largest Raman intensity, which is consistent with our experimental observation. In addition, the indistinguishable phonon at $\omega$ = 453 cm$^{-1}$, ascribed to E$_g$ symmetry, may be the noise in the spectrum, which appears to be missing since its intensity is very weak in Figure 3(a). To explore possible signatures of subtle spin-phonon coupling, the temperature-dependent phonon frequencies and linewidth (full width at half maximum), determined from fitting the peak to a Lorentzian, are plotted in Figure 3(b)-3(g). Under the assumption, that decay occurs to two phonons of frequencies $\omega_1$ and $\omega_2$ and three identical phonons of frequency $\omega/3$ [45-47], the phonon frequencies increase with decreasing temperature up to $T_C$ due to the anharmonic effect, which can be fitted by the equation [47]

$$\omega_{anh} = \omega_0 + A\left(1 + \frac{1}{e^{x_1}-1} + \frac{1}{e^{x_2}-1}\right) + B\left(1 + \frac{3}{e^{x_3}-1} + \frac{3}{(e^{x_3}-1)^2}\right), \quad (7)$$

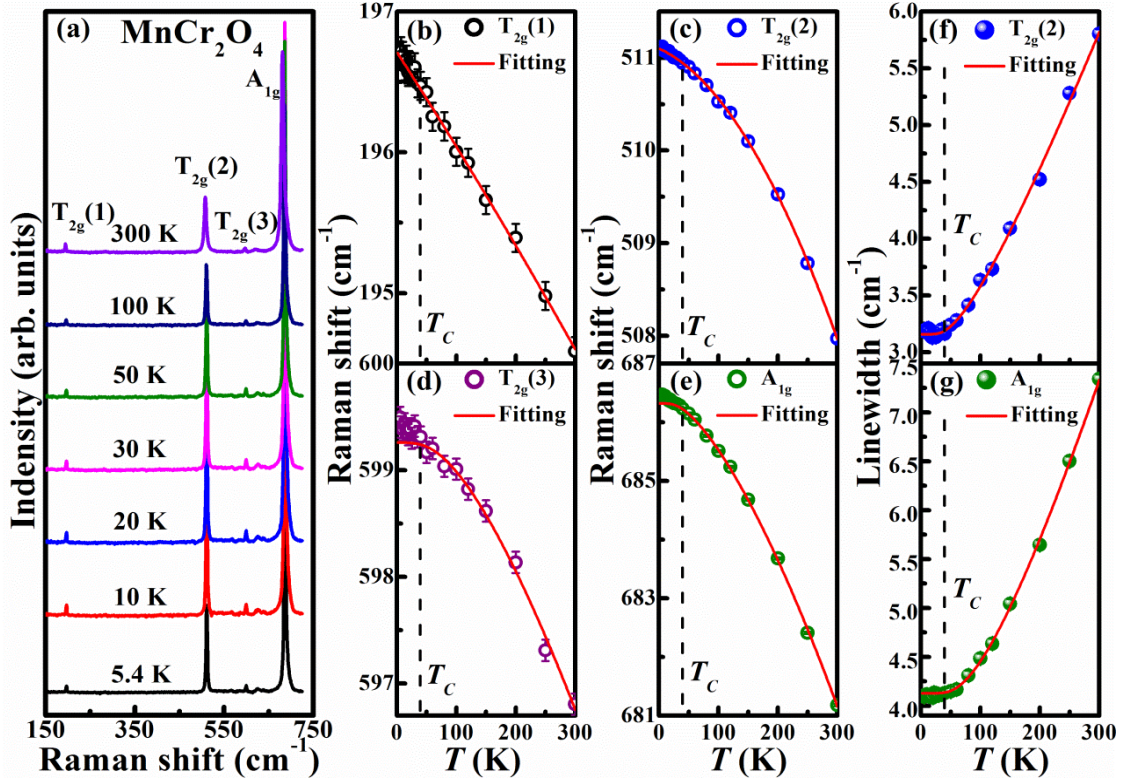

Figure 3. (a) Temperature-dependent Raman spectra of $MnCr_2O_4$; Temperature-dependent Raman phonon frequencies of (b) $T_{2g}(1)$, (c) $T_{2g}(2)$, (d) $T_{2g}(3)$, and (e) $A_{1g}$ modes. The red solid lines are the anharmonic contributions to the phonon frequencies fitted by Eq. (7); Temperature-dependent linewidth of (f) $T_{2g}(2)$, and (g) $A_{1g}$ modes.

where $\omega_0$, $A$ and $B$ are adjustable parameters, $x_1 = hc\omega_1/k_BT$, $x_2 = hc\omega_2/k_BT$, $x_3 = hc\omega/3k_BT$. $h$, $c$, $k_B$ and $T$ denote the Planck's constant, speed of light, Boltzmann's constant and temperature, respectively. This model describes adequately the temperature dependence of $T_{2g}(1)$ and $T_{2g}(2)$ phonon modes between 5.4 and 300 K, as shown in the Figure 3(b) and 3(c). The magnetic order below $T_C$, however, results in an very weak anomalous hardening of $T_{2g}(3)$ and $A_{1g}$ phonon modes, as evidenced in the Figure 3(d) and 3(e). It may be due to the fact that five modes ($A_{1g}$, $E_g$, $3T_{2g}$) of Raman-active have a weak response to the spin-lattice coupling. Therefore, the study of infrared spectroscopy is necessary in the future. Nevertheless, our results confirm that the cubic symmetry of the lattice is preserved even in the magnetically ordered ground state in $MnCr_2O_4$.

Figure 4(a) shows temperature-dependent electric polarization $P(T)$ and $M(T)$ along the [111] direction, in which the onset of pyroelectric current corresponds with the CSSO transition at $T_L$ = 17.4 K. Figure 4(b) displays how $P(T)$ depends on the different $H$ at $E_C$ = 0 V, showing a slight decreasing tendency of $T_L$ with increasing $H$. And the intensity of $P(T)$, tending to be saturated above about $H$ = 1 kOe, increases rapidly with increasing $H$. In Figure 4(c), one can observe that the field-dependent electric polarization $P(H)$ at 10 K and 17 K, respectively, in comparison with the $M(H)$ curve at 10 K. The synchronous reversal of the spontaneous $M$ and $P$ can be more directly confirmed by the continuous sweep field between +1.5 and -1.5 kOe at 10 K (see Figure 4(d)). The strikingly reversible and reproducible variation of the $P$ is observed without any noticeable decay in its magnitude. By combining this highly reproducible $P$ reversal with the ability to leave a permanent 'imprint' in the polarization with an applied magnetic field demonstrated in Figure 4(b) and 4(d), one can envision the low-field magnetoelectric effect has the promise of practical device applications, namely, a nonvolatile memory [48] where information is stored as electrically detectable and electrically controllable spin helicity.

As we know, a spontaneous electric polarization can appear when the spins form a transverse-spiral (cycloidal) modulation along a specific crystallographic direction and the spin rotation axis is not parallel to the propagation vector. The direction of the $P$ can be expressed by the equation [3,5,49-51]

$$\mathbf{P} = \gamma \sum_{\langle i,j \rangle} \mathbf{e}_{ij} \times (\mathbf{S}_i \times \mathbf{S}_j) \tag{8}$$

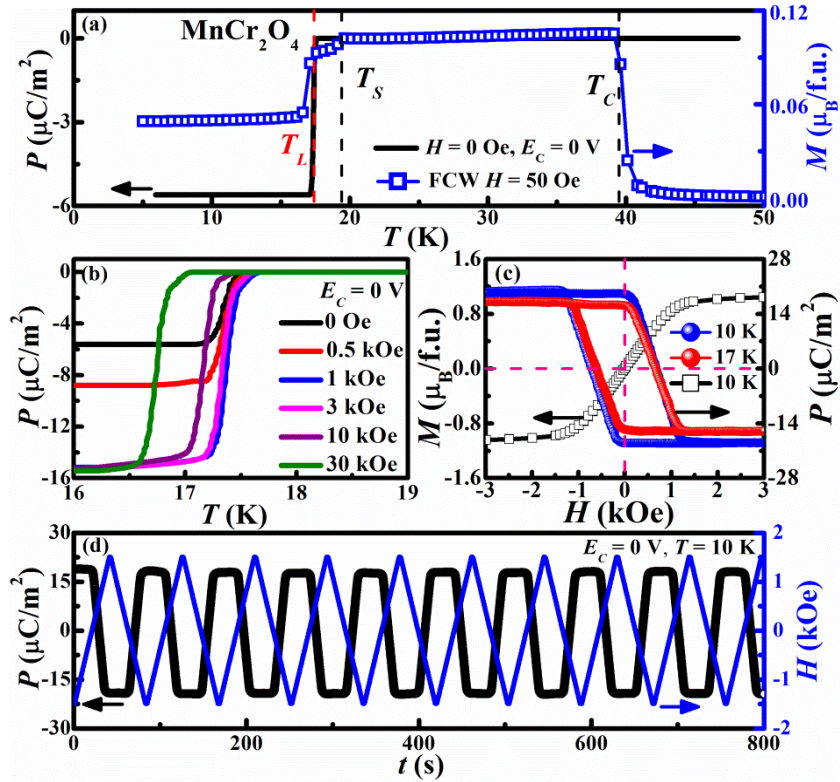

Figure 4. (Color online): (a) Temperature-dependent electric polarization $P(T)$ and $M(T)$ along the [111] direction; (b) $P(T)$ along the [111] direction around the $T_L$ with different applied field; (c) The $M(H)$ at 10 K, parallel to the <111> direction. The field-dependent electric polarization $P(H)$ at 10 K and 17 K, respectively, parallel to the <111> direction; (d) The synchronous reversal of the spontaneous M and P between +1.5 and -1.5 kOe at 10 K.

Here, $\gamma$ is a coefficient proportional to the spin-orbit coupling and super-exchange interactions as well as spin-lattice coupling, $\mathbf{e}_{ij}$ is along the propagation vector of a spiral structure, and $(\mathbf{S}_i \times \mathbf{S}_j)$ is parallel to the spin rotation axis. This model is termed the inverse Dzyaloshinskii–Moriya (DM) model or the spin-current model. The MF effect in many compounds, i.e., $R$MnO$_3$ ($R$ = Tb, Dy, etc.) [52,53], CoCr$_2$O$_4$ [1,2,16], and so on, can be explained by the Eq. (8). In addition, neutron scattering measurements have proved that the magnetic ground state of MnCr$_2$O$_4$ shows the coexistence of the conical spin order and LFIM, where the spontaneous magnetization vector is parallel to $[1\bar{1}0]$ or equivalent directions [21-24]. The conical spin order can be seen as composed of the cycloidal component and the ferrimagnetic component in this structure. Hence, as shown in Figure 5(b) and 5(c), the magnetoelectric effect of $MnCr_2O_4$ can also be categorized into cycloidal spiral origin based on the Eq. (8), which has been confirmed by the polycrystalline $MnCr_2O_4$ [15]. It should be mentioned that the spin rotation axis is parallel to the direction of ferrimagnetic component of the conical spin order, namely, $[1\bar{1}0]$ axis. Then, according to Eq. (8), the spontaneous polarization vector is expected to lie along the [001] axis. As mentioned above, we only test the magnetoelectric properties along the [111] direction, namely, the normal direction of easy-growth plane due to the eatremely small size in single-crystalline $MnCr_2O_4$. However, our results also verify that there is a strong magnetoelectric coupling in $MnCr_2O_4$. In addition, the ferroelectricity in $MnCr_2O_4$ still needs further investigation. This is due to the fact that the reverse electric field cannot reverse the direction of the *P* (see the Supplemental Material [32]).

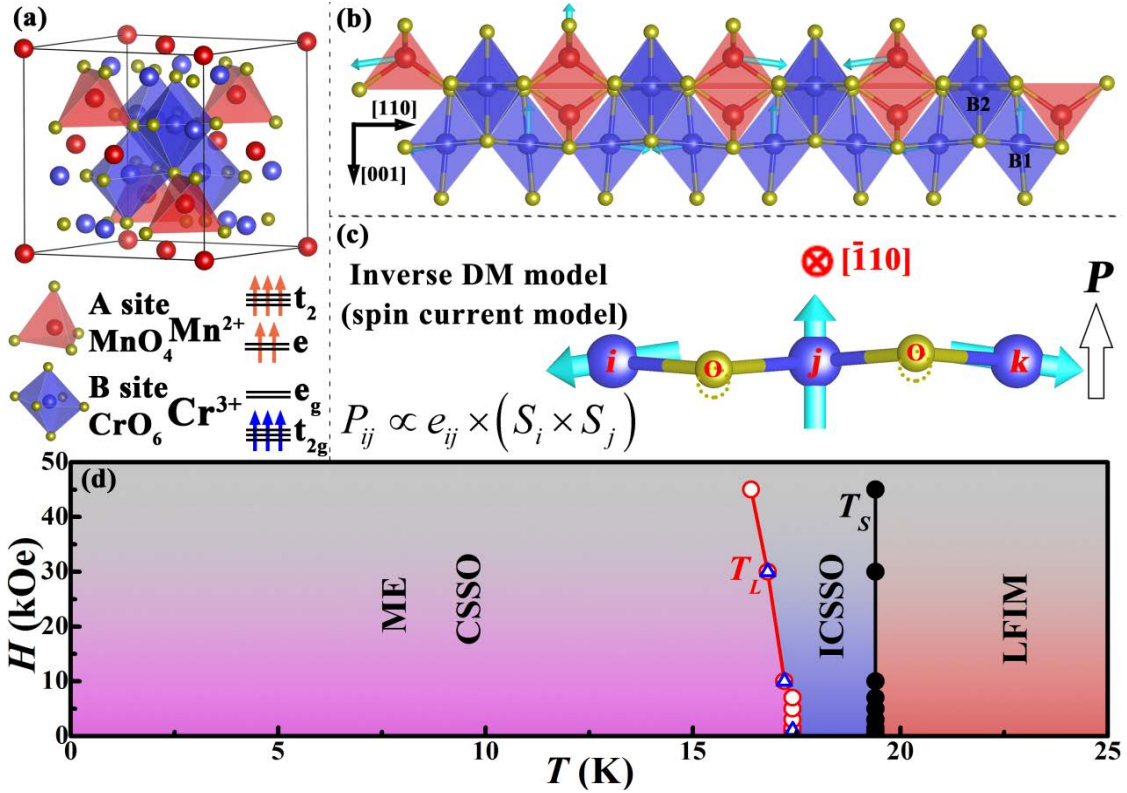

Figure 5. (Color online): (a) Schematic structure of spinel $MnCr_2O_4$. The electron configuration of $Mn^{2+}$ and $Cr^{3+}$ ions, located at the center of tetrahedral and octahedral $O^{2-}$ cages, respectively. Here, the splitting due to the local crystal field of the ions in the cubic phase and the effect of Hund coupling are only taken into account; (b) Low-temperature magnetic structure of spinel $MnCr_2O_4$ viewed along $[1\bar{1}0]$ direction; (c) The spin ($S_i$ and $S_j$) canting between the two sites (*i* and *j*) and the direction of induced polarization ***P***. (d) Schematic low-temperature phase diagram of $MnCr_2O_4$. The open circles and triangles refer to the phase-transition temperature obtained by ***M***(*T*) and ***P***(*T*) curves along the [111] directions, respectively.

## IV. Conclusion

Finally, based on the magnetization, SH, and magnetoelectric measurements, we plot the low-temperature phase diagram of $MnCr_2O_4$ (see Figure 5(d)). The first-order transition from the LFIM to the ICSSO at $T_S = 19.4$ K is almost field independent. The subsequent transition to the CSSO decreases with the external magnetic field at least up to 45 kOe, corresponding to the onset of spontaneous electric polarization with the external magnetic field. In addition, the temperature evolution of Raman spectrum between 5.4 and 300 K indicates that there is no structural phase change in $MnCr_2O_4$.

## Acknowledgements

This work was supported by the National Key Research and Development Program under contracts 2016YFA0401803 and 2016YFA0300404, the Joint Funds of the National Natural Science Foundation of China and the Chinese Academy of Sciences' Large-Scale Scientific Facility under contract U1432139 and U1532153, the National Natural Science Foundation of China under contract 11674326 and 11404339, Key Research Program of Frontier Sciences, CAS (QYZDB-SSW-SLH015), and the Nature Science Foundation of Anhui Province under contract 1508085ME103. This work used computational resources of the Texas Advanced Computing Center under Contract No.TG-DMR170070. H. L. Z. thanks the start-up funds from Arizona State University. The first author thanks Dr. C. Sun from University of Wisconsin-Madison for her assistance in editing the revised manuscript.